\documentclass[seceq]{ptptex}

\usepackage{graphicx}

\def\simge{\mathrel{%
       \rlap{\raise 0.511ex \hbox{$>$}}{\lower 0.511ex \hbox{$\sim$}}}}
\def\simle{\mathrel{
       \rlap{\raise 0.511ex \hbox{$<$}}{\lower 0.511ex \hbox{$\sim$}}}}

\title{
Lattice QCD thermodynamics with Wilson quarks%
}

\author{
Shinji \textsc{Ejiri}\footnote{ e-mail address:
ejiri@quark.phy.bnl.gov}  }

\inst{
Physics Department, Brookhaven National Laboratory, \\
Upton, New York 11973, USA 
}

\abst{
We review studies of QCD thermodynamics by lattice QCD simulations 
with dynamical Wilson quarks. 
After explaining the basic properties of QCD with Wilson quarks 
at finite temperature including the phase structure and the scaling 
properties around the chiral phase transition, we discuss the critical 
temperature, the equation of state and heavy-quark free energies. 
}

\begin{document}

\maketitle

\section{Introduction}

In order to remove theoretical uncertainties in the analysis of data 
from heavy-ion collision experiments, first principle calculations 
by the lattice QCD are indispensable and 
various interesting results have been already reported.
However, most lattice QCD studies at finite temperature $(T)$ 
and chemical potential $(\mu_q)$ have been performed using staggered-type 
quark actions with the fourth-root trick of the quark determinant so far.
Therefore, studies by the other actions such as Wilson-type quark 
actions are necessary to estimate systematic errors due to 
lattice discretization.

In this report, we want to highlight the QCD thermodynamics 
by numerical simulations with dynamical Wilson quarks. 
A systematic study of the QCD thermodynamics has been done 
by the CP-PACS Collaboration using the Iwasaki (RG) improved gauge action and 
the 2 flavor clover improved Wilson quark action several years 
ago\cite{cp1,cp2}.
Recently the WHOT-QCD Collaboration restarted the study by the 
same action adopting new technical developments\cite{whot1,whot2,whot3}.
In this review, we first explain in Sec.~\ref{sec:ps} the phase structure 
of 2 flavor QCD 
with Wilson quarks in the simulation parameter plane $(\beta, K)$. 
Then, the universality class of the chiral phase transition is discussed 
in Sec.~\ref{sec:o4}. 
An estimation of the critical temperature in the chiral limit is 
given in Sec.~\ref{sec:tc}. 
The equation of state (EoS) is discussed in Sec.~\ref{sec:eos}, 
highlighting fluctuations at finite density, related to the physics of 
the possible critical point in the $(T,\mu_q)$ plane., 
and also results of heavy-quark free energies are shown in 
Sec.~\ref{sec:fe}.

\section{Phase structure of QCD with Wilson quarks}
\label{sec:ps}

The lattice QCD with Wilson-type quarks is known 
to have a complicated phase structure due to the explicit violation of 
chiral symmetry and due to the existence of the parity-flavor broken 
phase (Aoki phase)\cite{aoki}. 
Therefore, a systematic study surveying a wide range of the parameter 
space is required to determine appropriate simulation parameters.
Moreover, lattice artifacts are large on coarse lattices used in most 
finite temperature simulations when the standard plaquette gauge 
action and the standard Wilson quark action are used, 
For example, unexpected strong phase transition is observed 
at intermediate quark masses in 2 flavor QCD.
Therefore, we have to improve the lattice action to reduce these 
lattice artifacts. 

\begin{figure}[tb]
  \begin{center}
    \begin{tabular}{cc}
      \includegraphics[width=59mm]{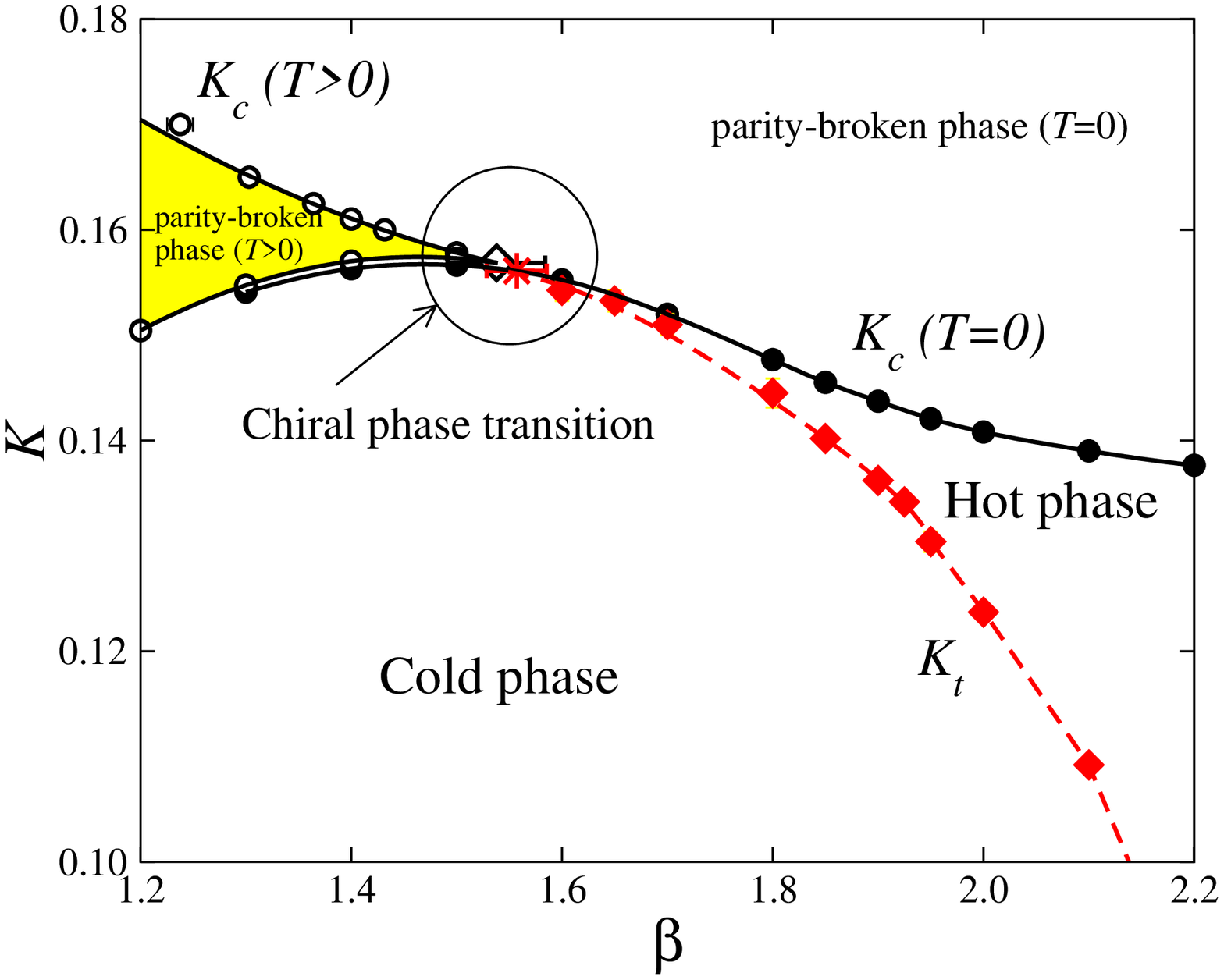} &
      \includegraphics[width=61mm]{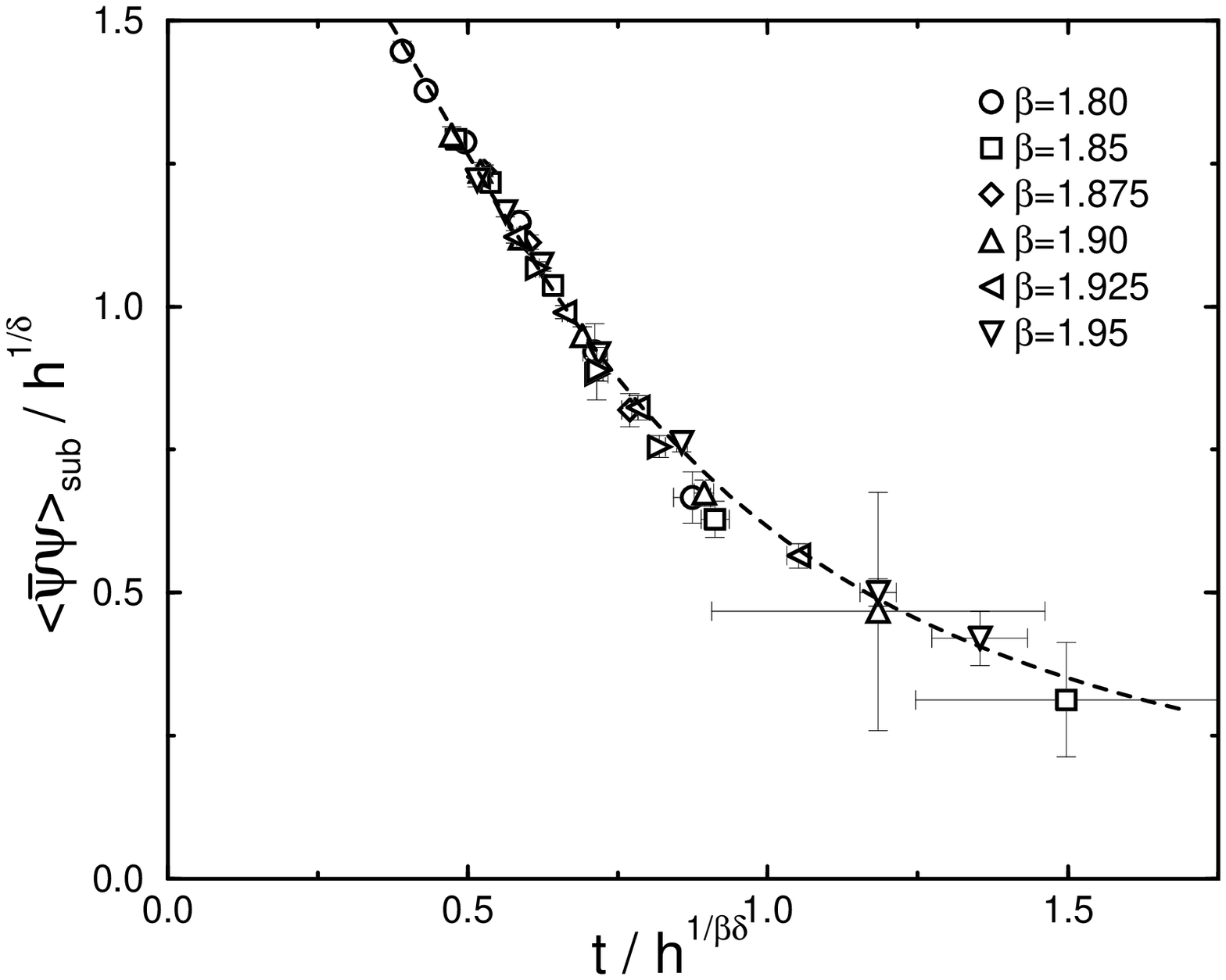}
    \end{tabular}
    \vspace*{-2mm}
    \caption{Left: Phase diagram for the RG-improved gauge and clover-improved  Wilson quark actions on an $Nt=4$ lattice.
    Right: O(4) scaling relation of the chiral condensate\cite{cp1}.
    }
    \label{fig1}
  \end{center}
\end{figure}

Figure~\ref{fig1} (left) is the phase diagram obtained by 
the CP-PACS Collaboration for 2 flavor QCD 
with the RG-improved gauge action combined with the clover-improved 
Wilson quark action\cite{cp1}.
The 2 flavor lattice QCD has two simulation parameters $K$ and $\beta$. 
$K$ is the hopping parameter in the quark action 
and $\beta$ is $6/g^2$ in the gauge action.
The solid line $K_c(T=0)$ is the location of the chilal limit. 
The pion mass decreases as $K$ increases from small $K$, 
and vanishes on the line $K_c(T=0)$ at zero temperature. 
In the region above $K_c(T=0)$, a parity-flavor symmetry 
of the Wilson-type quark action is broken spontaneously\cite{aoki}. 
At zero temperature, the boundary of the parity-flavor broken phase 
is known to form a sharp cusp touching the free 
massless fermion point $K=1/8$ at $\beta=\infty$.
However, the region above $K_c(T=0)$ is an unphysical parameter space, 
and we usually perform simulations below $K_c$. 
The region below $K_c(T=0)$ corresponds to the physical QCD with 
$1/K - 1/K_c$ being proportional to the quark mass $m_q$. 

At finite temperature, the parity-flavor broken phase retracts 
from the large $\beta$ region.
The colored region with the boundary $K_c(T>0)$ in Fig.~\ref{fig1} (left) 
is the parity-flavor broken phase at finite temperature for 
the temporal lattice size $N_t=4$. 
When $N_t$ is fixed, 
the temperature $T=(N_t a)^{-1}$ becomes higher as $\beta$ is increased, 
since the lattice spacing $a$ becomes smaller.
The dashed line $K_t$ is the pseudo-critical line separating 
the hot and cold phases for $N_t=4$.
The region to the right of $K_t$ (larger $\beta$) is the high 
temperature quark-gluon plasma (QGP) phase, and that to the left 
(smaller $\beta$) is the low temperature hadron phase. 
The crossing point of the $K_c(T=0)$ and the $K_t$ is the chiral phase 
transition point. 

As shown in this figure, the line that the pion mass vanishes 
at $T>0$ ($K_c(T>0)$) runs along the line of the chiral limit 
in the low temperature phase,
while the $K_c(T>0)$ line bends sharply at the chiral phase 
transition point and goes to the unphysical region above $K_c(T=0)$.
This is consistent with the picture that the massless pion, i.e. 
the Goldstone boson associated with spontaneous chiral symmetry breaking,
appears only in the cold phase. 

\section{O(4) scaling relation}
\label{sec:o4}

Next, we discuss the natures of the chiral phase transition 
in the critical region. 
The order of the phase transition is expected to be second order for 
2 flavor QCD and first order for 3 flavor QCD in the chiral limit.
The confirmation of this expectation is an important step toward 
the clarification of the QCD transition in the real world. 
In this section, we discuss the scaling behavior of 2 flavor QCD.

When the chiral transition of 2 flavor QCD is second order, 
the transition is expected to be in the same universal class with 
a 3-dimensional O(4) spin model. 
With the identifications
$M \sim \langle \bar{\Psi} \Psi \rangle$ for the magnetization, 
$h \sim m_{q} a$ for the external magnetic field
and $t \sim \beta - \beta_{ct}$ for the reduced temperature,
where $\beta_{ct}$ is the chiral transition point, 
we expect the same scaling behavior as the O(4) spin model. 

The O(4) scaling was first tested with staggered fermion by Karsch and 
Laermann\cite{Kar94}. 
The study was extended to a wider range of the quark mass and 
lattice sizes by several groups \cite{JL98,Bi98,Ber00,pisa05}.
However, an agreement of the critical exponents between the O(4) spin model 
and QCD with 2 flavors of staggered quarks has not been obtained.

For the case of Wilson quarks, 
Iwasaki et al.\cite{o4scale} investigated the scaling relation, 
\begin{eqnarray}
M / h^{1/\delta} = f (t / h^{1/\beta \delta}), \label{eq:scfn}
\end{eqnarray}
using the standard Wilson quark action coupled to RG-improved gluon.
They identified the subtracted chiral condensate 
defined by an axial Ward-Takahashi identity\cite{bochicchio},
$\langle \bar{\Psi} \Psi \rangle_{\rm sub}= 2m_q a 
  Z \sum_{x} \langle \pi (x) \pi(0) \rangle, $
as the magnetization of the spin model.
Here, the quark mass $m_q$ is defined by an axial vector 
Ward-Takahashi identity\cite{bochicchio}, $m_q^{\rm AWI}$,
and the tree-level renormalization coefficient $Z = (2K)^2$ was adopted.
They found that the scaling relation Eq.~(\ref{eq:scfn}) is well satisfied 
with the critical exponents and the scaling function of the O(4) spin model. 

The O(4) scaling has been obtained also with improved Wilson quarks.
Figure~\ref{fig1} (right) is the result for the case of 
the clover-improved Wilson quark action coupled with the RG gauge action, 
obtained on a $16^3 \times 4$ lattice\cite{cp1}. 
The vertical axis is $M/h^{1/\delta}$ and the horizontal axis is 
$t/h^{1/\beta \delta}$, 
where $\beta$ and $\delta$ are the critical exponents obtained 
in the O(4) spin model. 
The dashed line is the O(4) scaling function. 
They fitted the data to the scaling function adjusting $\beta_{ct}$ and 
the scales of two axes.
As seen from Fig.~\ref{fig1} (right), 
QCD data is well described by the O(4) scaling ansatz.
This result suggests that the chiral phase transition is of second order 
for 2 flavor QCD.

\begin{figure}[tb]
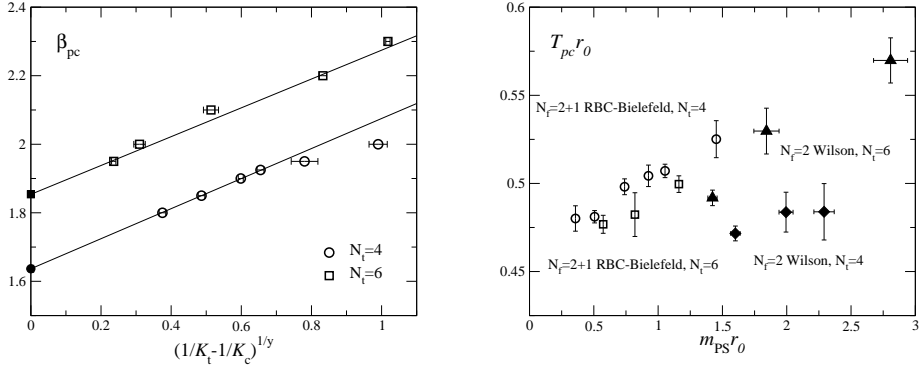

  \begin{center}
    \begin{tabular}{cc}
      \includegraphics[width=56mm]{fig/invKhvb_QM06.eps} &
     \hspace{5mm}
      \includegraphics[width=56mm]{fig/tcr0vsmpir0_QM06.eps}
    \end{tabular}
    \vspace*{-2mm}
    \caption{Left: The pseudo-critical point $\beta_{pc}$
    as a function of $m_q \sim 1/K - 1/K_c$ for $N_t=4$ (circle) 
    and $N_t=6$ (square)\cite{whot2}.  
    Right: Comparison of $T_{pc}$ scaled by $r_0$ between 
    the staggered quark action (open symbol) \cite{Che06} and 
    the Wilson quark action (filled symbol) \cite{whot2} for $N_t=4$ and 6.
    }
    \label{fig2}
  \end{center}
\end{figure}

\section{Critical temperature of 2 flavor QCD}
\label{sec:tc}

The critical temperature ($T_c$) is one of the most fundamental quantities 
in the QCD thermodynamics and is important in phenomenological studies of 
heavy ion collisions.
Recently several groups \cite{Ber05,Che06,Aoki06}
have tried to determine $T_c$ near the physical mass parameter 
in 2+1 flavor QCD by simulations with improved staggered quarks. However, 
the results are still contradictory to each other. 

The WHOT-QCD Collaboration reported recently a preliminary result of $T_c$ 
for 2 flavor QCD renewing the analysis done in Ref.~\citen{cp1} \cite{whot2}.
They determined the pseudo-critical points $\beta_{pc}$ defined from 
the peak of the Polyakov loop susceptibility on 
$16^3 \times 4$ and $16^3 \times 6$ lattices, 
as a function of the hopping parameter $K$.

As seen in the previous section, the subtracted chiral condensate 
satisfies the scaling behavior with the critical exponents and 
scaling function of the 3-dimensional O(4) spin model.
Assuming that the pseudo-critical temperature
from the Polyakov loop susceptibility
follows the same scaling
law as the O(4) spin model, i.e.  $t_{pc} \sim h^y$
with $y \equiv \beta \delta = 0.537(7)$, 
the data of $\beta_{pc}$ in Fig.~\ref{fig2} (left) are fitted by
$\beta_{pc}=\beta_{ct}+Ah^{1/y}$
with two free parameters, $\beta_{ct}$ and $A$, in the range of 
$\beta=1.8$--1.95 for $N_t=4$ and $\beta=1.95$--2.10 for $N_t=6$.
First, they adopted the definition $m_q a \sim 1/K - 1/K_c$ 
as the quark mass where $K_c$ is the chiral point where the pion 
mass vanishes at $T=0$ for each $\beta$. 
The critical temperature $T_c$ is calculated in the chiral limit 
using $T=1/(N_t a)$. The lattice spacing $a$ is estimated from the vector 
meson mass assuming $m_{\rm V}(T=0)=m_{\rho}=770$ MeV at $\beta_{ct}$ on $K_c$. 
By this procedure,
They obtained preliminary results of $T_c = 183(3)$ MeV for $N_t=4$ and 
174(5) MeV for $N_t=6$. 
They also calculated $\beta_{ct}$ using the relation of 
$m_q^{\rm AWI} \propto m_{\rm PS}^2$, 
where $m_{\rm PS}$ is the pseudo-scalar meson mass  and $m_q^{\rm AWI}$ is
 the quark mass obtained from the axial vector Ward-Takahashi identity. 
The results of $T_c$ are 173(3) MeV $(N_t=4)$, 167(3) MeV $(N_t=6)$ for 
$h=(m_{\rm PS} a)^2$ and 176(3) MeV $(N_t=4)$ for $h=m_q^{\rm AWI} a$.
It is noted that these O(4) fits reproduce the data of $\beta_{pc}$ much better 
than a linear fit $\beta_{pc}=\beta_{ct}+Ah$. 
A tentatively conclusion is that
the critical temperature in the chiral limit is in 
the range 170--186 MeV for $N_t=4$ and 164--179 MeV for $N_t=6$. 
There is still a large uncertainty from the choice of the fit ansatz. 
To remove this, further simulations at lighter quark masses are necessary.

Next, we compare these results with those of a staggered quark action.
We plot the results of the pseudo-critical temperature ($T_{pc}$) 
in unit of Sommer scale $(r_0)$ as a 
function of $m_{\rm PS} r_0$ in Fig.~\ref{fig2} (right) together with 
those by the RBC-Bielefeld Collaboration using 2+1 flavor p4-improved staggered 
quark action \cite{Che06}. 
As seen in this figure, results of $T_{pc}$ obtained by different quark 
actions seem to approach the same function of $m_{\rm PS} r_0$ 
as $N_t$ increases.

\begin{figure}[tb]
  \begin{center}
    \begin{tabular}{cc}
      \includegraphics[width=63mm]{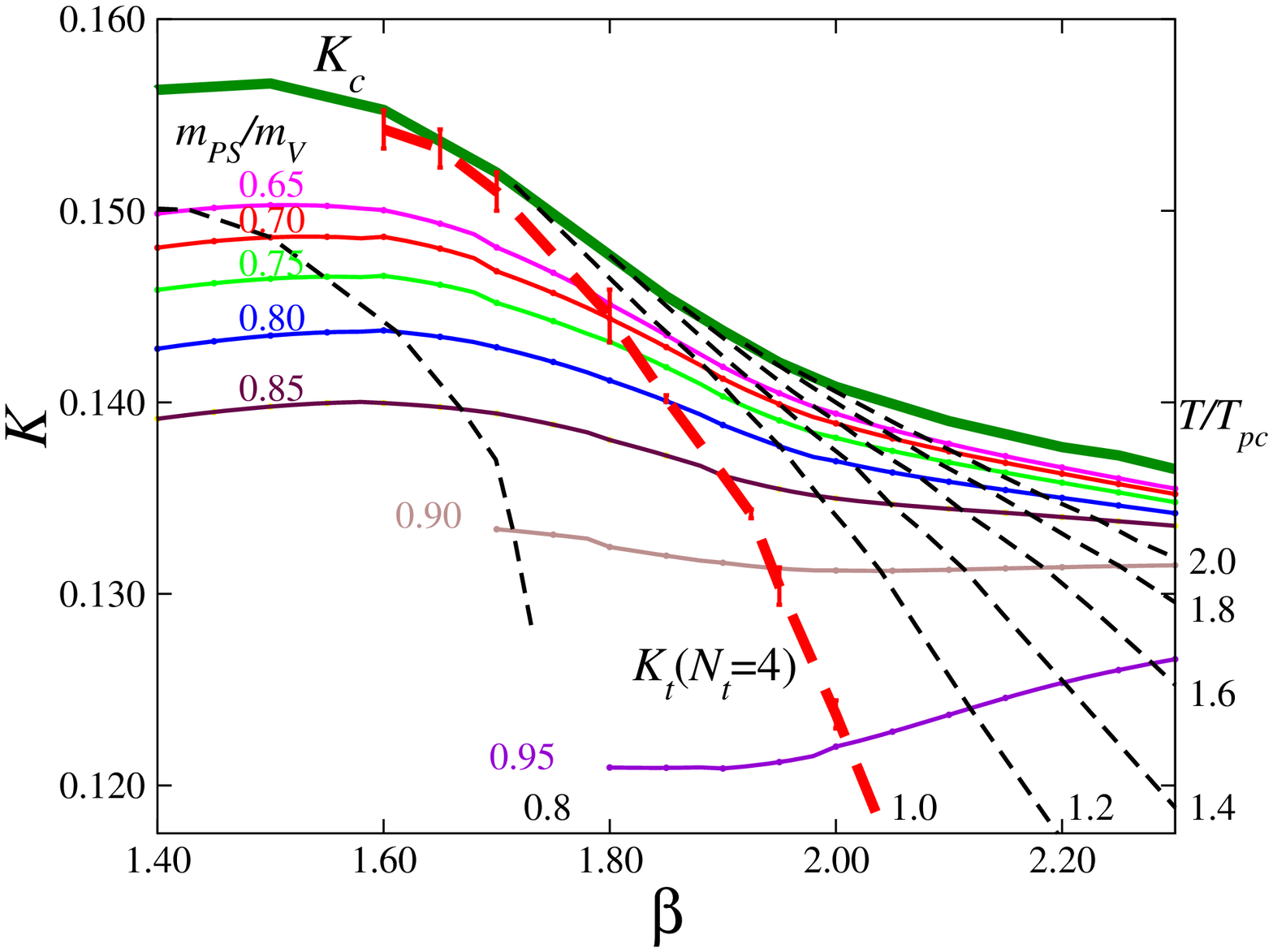} &
      \includegraphics[width=66mm]{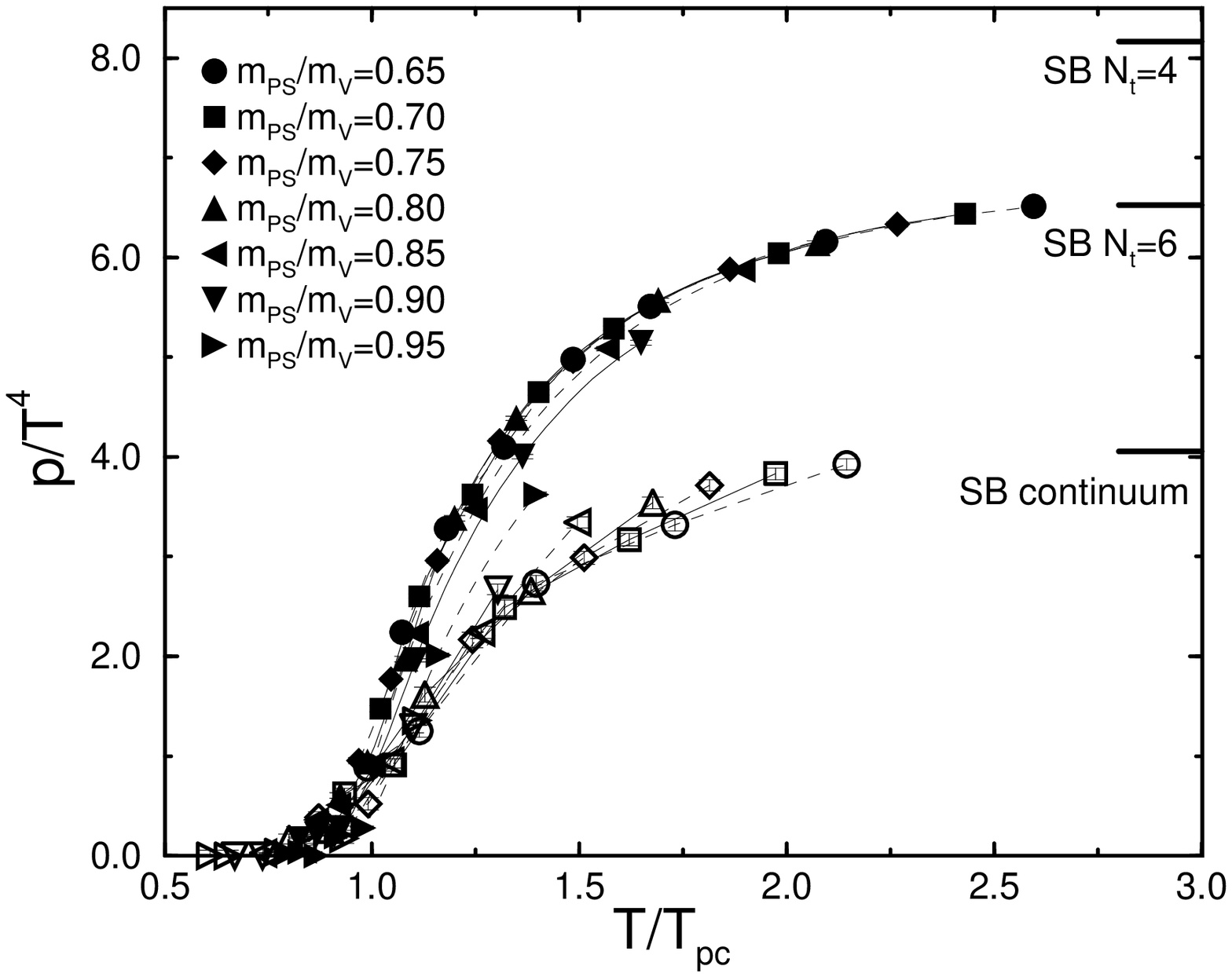}
    \end{tabular}
    \vspace*{-2mm}
    \caption{Left: Lines of constant $m_{PS}/m_V$ (solid line) and  
    $T/T_{pc}$ (dashed line) on an $Nt=4$ lattice.
    Right: Pressure as a function of $T/T_{pc}$ on an $Nt=4$ (fulled) 
    and 6 (open) for each $m_{PS}/m_V$\cite{cp2}.
    }
    \label{fig3}
  \end{center}
\end{figure}

\section{Equation of state at $\mu_q=0$ and $\mu_q \neq 0$}
\label{sec:eos}

The studies of the equation of state (EoS) can provide basic input 
for the analysis of the experimental signatures for QGP formation, e.g. 
the EoS will control the properties of any hydrodynamic expansion. 

For the study at zero chemical potential, 
the integral method is commonly used. 
This method is based on the equation for pressure, $p= (T/V) {\rm ln} Z$, 
where $Z$ is the partition function.
Because the derivatives of the partition function can be expressed 
by expectation values of operators, 
which are computable by a Monte-Carlo simulation, 
we obtain the pressure by integrating this expectation value in the 
parameter space.
For the case of the Wilson quark, we have 
\begin{eqnarray} 
\frac{p}{T^4} = - N_t^4 
\int^{(\beta,K)} {\rm d} \xi \left\{ \frac{1}{N_{s}^{3} N_{t}}
\left\langle \frac{\partial S}{\partial \xi} \right\rangle 
- ({\rm value \ at \ } T=0) 
\right\}
\label{eq:int}
\end{eqnarray}
with ${\rm d} \xi=({\rm d} \beta', {\rm d} K')$ on the integration path.
The starting point of the integration path should be chosen such that 
$p \approx 0$ there.

The CP-PACS Collaboration carried out a systematic calculation of EOS 
with a 2 flavor Wilson-type quark action at $\mu_q=0$. 
They used the RG gauge and the clover quark actions with $16^3 \times 4$ 
and $16^3 \times 6$ lattices\cite{cp2}. 
The translation from the results obtained by Eq.~(\ref{eq:int}) as functions 
of $(\beta, K)$ to those of physical parameters can be done using 
Fig.~\ref{fig3} (left). 
The thin solid lines shows the lines of 
constant physics (LCP), which they determine by $m_{\rm PS}/m_{\rm V}$ 
(the ratio of pseudo-scalar and vector meson masses at $T=0$). 
The chiral limit $K_c$ corresponds to LCP for $m_{\rm PS}/m_{\rm V}=0$.
The bold dashed line denoted as $K_t(N_t=4)$ represents the
pseudo-critical line $T/T_{pc}=1$ at $N_t=4$.
The thin dashed lines represent the lines of constant $T/T_{pc}$ 
estimated by $T/m_V=(N_t m_V a)^{-1}$.

The right panel of Fig.~\ref{fig3} is the result of the pressure as 
a function of temperature for each LCP.
Filled and open symbols are the results for $N_t=4$ and 6 respectively. 
Different shapes of the symbol correspond to different values of 
$m_{PS} / m_{V}$, i.e. different quark masses. 
This figure shows that the pressure is almost independent of the quark mass 
in a wide range of $m_{PS}/m_{V}$.
However, the $N_t$-dependence is sizeable, 
hence further simulations with large $N_t$ are important.

\begin{figure}[tb]
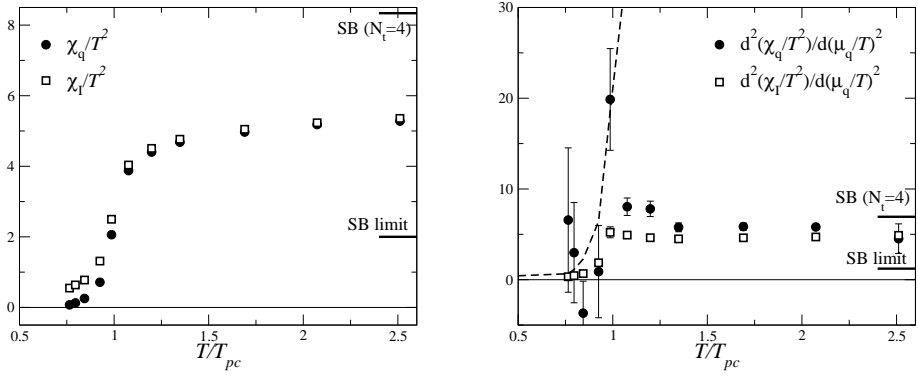

  \begin{center}
    \begin{tabular}{cc}
      \includegraphics[width=56mm]{fig/qnsm080c2_qm06.eps} &
      \hspace*{5mm}
      \includegraphics[width=56mm]{fig/qnsm080c4_qm06.eps} 
    \end{tabular}
    \vspace*{-2mm}
    \caption{Left: Quark number (circle) and isospin (square)
    susceptibilities at $\mu_q = \mu_I = 0$.
    Right: The second derivatives of these susceptibilities.
    }
    \label{fig4}
  \end{center}
\end{figure}

On the other hand, the studies of the EoS at non-zero baryon number density 
is attracting interest widely since the chemical potential dependence 
of the EoS should explain the difference of experimental results 
obtained at different beam energy. 
Moreover, hadronic fluctuations at finite densities are closely 
related to the appearance of the critical point in the $(T, \mu_q)$ plane 
and may be experimentally testable by an event-by-event analysis of 
heavy ion collisions. 
The fluctuations can also be studied by numerical simulations of 
lattice QCD calculating the quark number and isospin susceptibilities,  
$\chi_q$ and $\chi_I$. They correspond to
the second derivatives of the pressure with respect to $\mu_q$ 
and $\mu_I$, where $\mu_I$ is the isospin chemical potential. 
From a phenomenological argument in the sigma model, 
$\chi_q$ is singular at the critical point, 
whereas $\chi_I$ shows no singularity there. 

To investigate the EoS at finite density, the Bielefeld-Swansea 
Collaboration proposed 
a Taylor expansion method \cite{BS03}. Since Monte-Carlo simulations 
cannot be performed at finite density due to the sign problem, 
one evaluates the $\mu_q$ dependence of ${\rm ln}Z$ by computing 
the higher order derivatives of ${\rm ln} Z$ instead 
of integrating the first derivative.

The WHOT-QCD Collaboration performed simulations at 
$m_{\rm PS} / m_{\rm V} = 0.65$ and 0.80  
on a $16^3 \times 4$ lattice with the improved Wilson quark action.
They calculated the second and forth derivatives of pressure which correspond to 
the $\chi_q$ and $\chi_I$ and their second derivatives  
with respect to $\mu_q$ and $\mu_I$ 
at $\mu_q = \mu_I =0$. (Note that the odd derivatives are zero at $\mu_q=0$.) 

The left panel of Fig.~\ref{fig4} shows 
$\chi_q/T^2$ (circle) and $\chi_I/T^2$ (square) 
at $m_{\rm PS}/m_{\rm V}=0.8$ and $\mu_q = \mu_I =0$ as functions of $T/T_{pc}$.
It is found that $\chi_q/T^2$ and $\chi_I/T^2$ 
increase sharply at $T_{pc}$, in accordance with the expectation 
that the fluctuations in the QGP phase are much larger 
than those in the hadron phase.
Their second derivatives $\partial^2 (\chi_q/T^2)/ \partial (\mu_q/T)^2$ 
and $\partial^2 (\chi_I/T^2)/ \partial (\mu_q/T)^2$ are shown 
in Fig.~\ref{fig4} (right). 
The basic features are quite similar to those found previously with the p4-improved staggered fermions \cite{BS03}. 
$\partial^2 (\chi_I/T^2)/ \partial (\mu_q/T)^2$ remains small around $T_{pc}$,
suggesting that there are no singularities in $\chi_I$ at non-zero density. 
On the other hand, we expect a large enhancement in the quark number 
fluctuations near $T_{pc}$ as approaching the critical point in the 
$(T, \mu_q)$ plane. 
The dashed line in Fig.~\ref{fig4} (right) 
is a prediction from the hadron resonance gas model at low temperature,
$\partial^2 \chi_q/ \partial \mu_q^2 \approx 9 \chi_q/T^2$.
Although  current statistical errors 
in Fig.~\ref{fig4} (right) are still large, 
$\partial^2(\chi_q/T^2) / \partial (\mu_q/T)^2$ near $T_{pc}$ is 
much larger than that at high temperatures.
At the right end of the figure, values 
of free quark-gluon gas (Stefan-Boltzmann gas) 
for $N_t=4$ and for $N_t = \infty$ limit are shown.
Since the lattice discretization error in the EoS is 
known to be large at $N_t=4$ with their quark action,
it is needed to extend this study to larger $N_t$ for the continuum extrapolation.

\section{Heavy quark free energies}
\label{sec:fe}

Finally, we discuss a free energy between static quarks.
Clarification of the interaction between heavy quarks in QGP is important 
to understand the properties of charmoniums in heavy ion collisions. 
The heavy quark free energy $F_M$ in various color channels $M$ 
can be measured separately on the lattice by the correlations of 
the Polyakov loop with an appropriate gauge fixing. 

Recently, the heavy quark free energy for 2 flavor QCD with dynamical 
Wilson quarks in the Coulomb gauge are studied by the WHOT-QCD Collaboration, 
using the same configuration for the calculation of the EoS 
in the previous section\cite{whot3}. With the improved actions they adopted,
the rotational symmetry is well restored in the heavy quark free
energies\cite{cp3}, hence it is not necessary to introduce terms correcting 
lattice artifacts at short distances to analyze the data.

\begin{figure}[tbp]
  \begin{center}
    \begin{tabular}{cc}
      \resizebox{64mm}{!}{\includegraphics{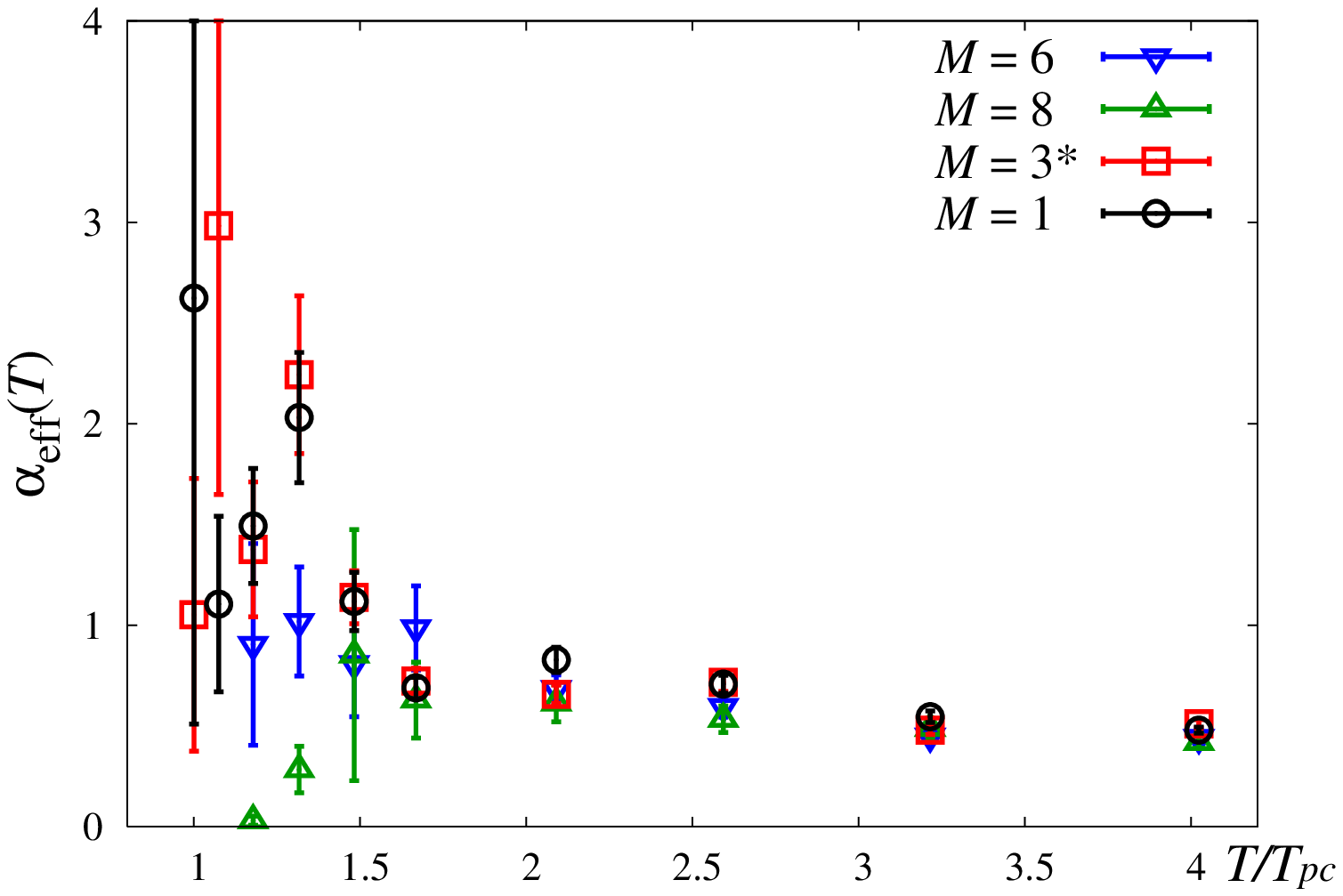}} &
      \resizebox{64mm}{!}{\includegraphics{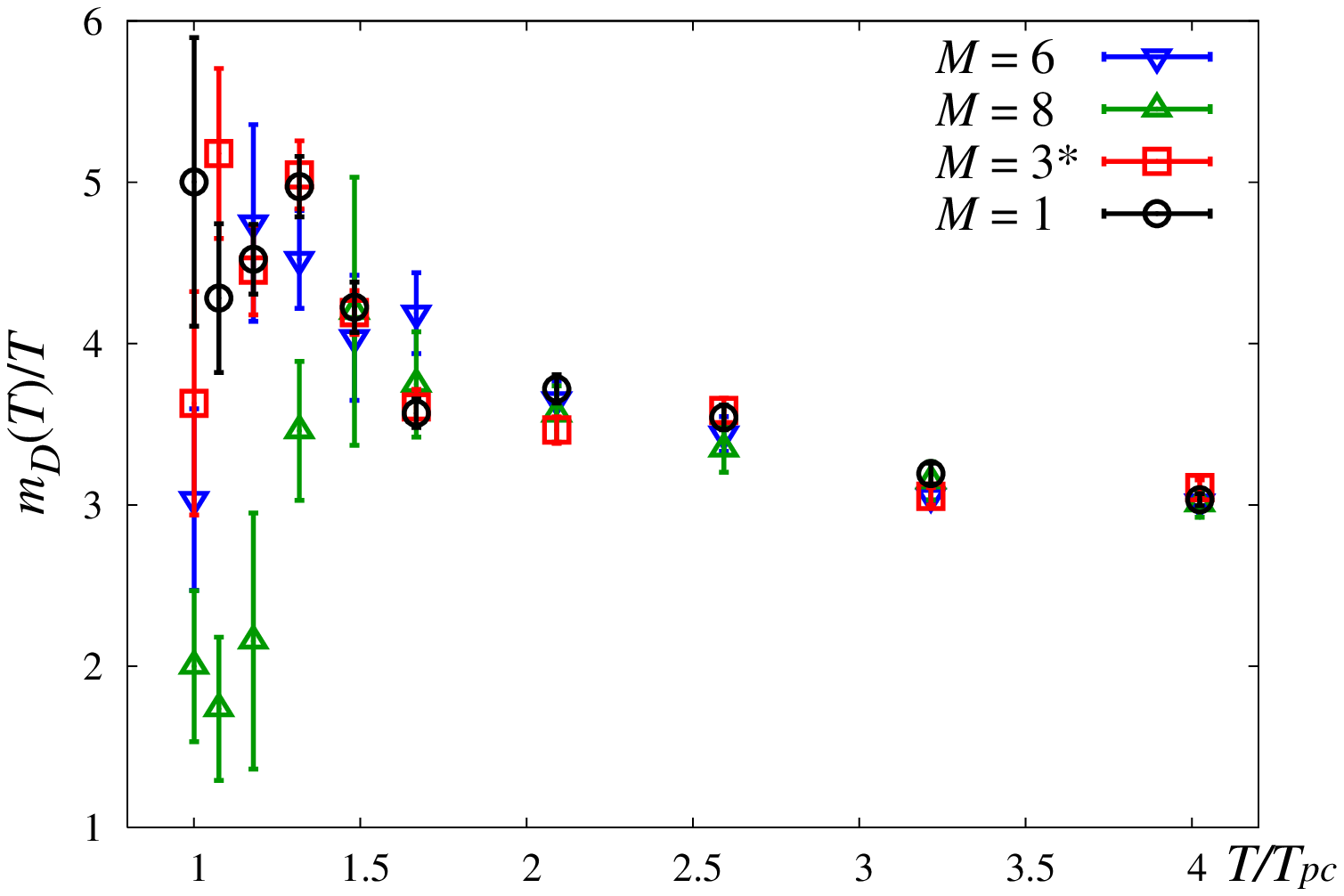}} 
    \end{tabular}
    \caption{The effective running coupling $\alpha_{\rm eff}(T)$ (left) and
    Debye screening mass $m_D(T)$ (right) for each color channel
    as a function of temperature at $m_{\rm PS}/m_{\rm V} = 0.65$.
    }
    \label{fig5}
  \end{center}
\end{figure}

They found that, at $T > T_{pc}$,
the free energies of $QQ$ and $Q \overline{Q}$ 
normalized to be zero at large separation show attraction (repulsion)
in the color singlet and anti-triplet channels (color octet and sextet
channels), and fitted the free energy data in each channel by 
the screened Coulomb form, 
\begin{equation}
F_M(r,T) - F_M(\infty,T) = C(M) \frac{\alpha_{\rm eff}(T)}{r} e^{-m_D(T) r} ,
\label{eq:SCP}
\end{equation}
where $\alpha_{\rm eff}(T)$ and $m_D(T)$ 
are the effective running coupling and Debye screening mass, respectively.
The Casimir factor $C(M) \equiv \langle \sum_{a=1}^{8} t_1^a\cdot t_2^a \rangle_M$ 
for color channel $M$  is explicitly given  by 
$
C({\bf 1})  = -\frac{4}{3}, 
C({\bf 8})  =  \frac{1}{6}, 
C({\bf 6})  =  \frac{1}{3}, 
C({\bf 3}^*)= -\frac{2}{3}.
$
The results of $\alpha_{\rm eff}(T)$ and $m_D(T)$ are shown
in Fig.~\ref{fig5} for $m_{\rm PS}/m_{\rm V} = 0.65$.
They found that there is no significant channel dependence in  
 $\alpha_{\rm eff}(T)$ and $m_D(T)$
at sufficiently high temperatures $(T \simge 2T_{pc})$.
In other words, the channel dependence in the free energy
can be well absorbed in the kinematical Casimir factor at high temperatures, 

The magnitude and the $T$-dependence of $m_D(T)$ is consistent 
with the next-to-leading order calculation in thermal perturbation theory.
Moreover it is also well approximated by the leading order form
with an ``effective'' running coupling defined 
from  $\alpha_{\rm eff}(T)$.
On the other hand, by comparing these results with the results by 
an improved staggered quark action \cite{Kac05}, 
they found that $\alpha_{\rm eff}(T)$ 
does not show appreciable difference
while $m_D(T)$ in the Wilson quark action is 
larger than that of the staggered quark action by 20\%.
To draw a definite conclusion, however, simulations with smaller
lattice spacings, i.e., larger lattice sizes in the temporal direction
(such as $N_t=6$ or larger) at smaller quark masses are required.

\section{Sumarry}
\label{sec:con}

We reported the current status of the study of QCD thermodynamics using a 
Wilson quark action. 
The basic properties of QCD with 2 flavors of Wilson quarks such as the 
phase structure and the scaling behavior have been already studied. 
Moreover, the qualitative agreements between the results obtained 
by a Wilson quark action and a staggered quark action have been 
discussed for some thermodynamic quantities.
The quantitative improvements are important in the future studies. 

\vspace{3mm}
I would like to thank the members of the WHOT-QCD Collaboration for 
collaboration, discussions and comments.
This work is supported by Grants-in-Aid of the Japanese 
MEXT (No.~18740134) and Sumitomo Foundation (No.~050408).

\end{document}